\documentclass[12pt]{article}
\usepackage{amsfonts,amsmath,amssymb,epsf}
\usepackage{graphicx,color}
\newcommand{\al}{\alpha'}
\newcommand{\de}{\partial}
\newcommand{\be}{\begin{equation}}
\newcommand{\ba}{\begin{eqnarray}}
\newcommand{\ea}{\end{eqnarray}}
\newcommand{\ee}{\end{equation}}

\newcommand{\f}{\frac}
\newcommand{\s}{\sqrt}

\newcommand{\ti}{\tilde}
\newcommand{\ap}{\alpha}

\newcommand{\ddd}{\cdot\cdot\cdot}
\newcommand{\no}{\nonumber \\}

\newcommand{\ov}{\overline}

\begin{document}

\begin{titlepage}
\thispagestyle{empty}

KUNS-2133, UTAP-595, RESCEU-6/08,  IPMU-08-0016
\begin{flushright}
\end{flushright}

\bigskip

\begin{center}
\noindent{\Large \textbf{Decaying D-branes and Moving Mirrors}}\\
\vspace{2cm} \noindent{ Tomoyoshi
Hirata$^a$\footnote{e-mail:hirata@gauge.scphys.kyoto-u.ac.jp},
  Shinji
  Mukohyama$^{b,c}$\footnote{e-mail:mukoyama@utap.phys.s.u-tokyo.ac.jp}
 and Tadashi
Takayanagi$^a$\footnote{e-mail:takayana@gauge.scphys.kyoto-u.ac.jp}}\\
\vspace{1cm}

 {\it $^a$ Department of Physics, Kyoto University, Kyoto 606-8502, Japan
 }

{\it $^b$ Department of Physics and Research Center for the Early
 Universe, University of Tokyo, Tokyo, 113-0033, Japan
 }

{\it $^c$  Institute for the Physics and Mathematics of the Universe,
 University of Tokyo, Chiba 277-8568, Japan
}

\vskip 2em
\end{center}

\begin{abstract}
We present an exact time-dependent solution to the effective D-brane
world-volume theory which describes an inhomogeneous decay of a
brane-antibrane system. We compute the quantum energy flux induced
by the particle creation in this inhomogeneous and time-dependent
background. We find that this calculation is essentially equivalent
to that of the moving mirror system. In the initial stage, the
energy flux turns out to be thermal with the temperature given by
the inverse of the distance between the brane and the antibrane.
Later it changes its sign and becomes a negative energy flux. Our
result may be relevant for the (p)reheating process or/and the
evolution of cosmic string network after stringy brane inflation.
\end{abstract}

\end{titlepage}


\section{Introduction}
\setcounter{equation}{0} {\hspace{5mm} The brane-antibrane system
\cite{Sen:1998sm} has been studied intensively until recently as a
typical example of unstable non-supersymmetric systems in string
theory \cite{Sen}. When a brane and an antibrane are close to each
other, the open string spectrum includes tachyons. The condensation
of open string tachyons offers us an important time-dependent
background.

When the distance between a brane and an antibrane is larger than the
string scale, there exits no tachyon. However, by an instanton
effect or a small dynamical perturbation induced by collisions with
other objects, it is possible that the distance between them becomes
the string scale at a particular region or a point. In this setup, we
expect that there exists a localized open string tachyon and the
tachyon condensation occurs only in that region. This leads to a
recombination of the brane and antibrane. Once this happens, this
system annihilates via the time evolution and eventually the whole
system will disappear radiating closed strings. However, the study
in this direction has been rather few even at present.

The brane-antibrane system plays important roles not only in
theoretical foundations of string theory but also in its
cosmological applications. (See ref.~\cite{McAllister:2007bg} and
references therein for a recent review of string cosmology.) Indeed,
a string theoretic realization of hybrid inflation, called warped
brane inflation~\cite{Kachru:2003sx}, was made possible by
non-trivial dynamics of the brane-antibrane system. In this
scenario, the distance between a brane and an antibrane embedded in
a warped geometry plays the role of inflaton. Our $4$-dimensional
universe parallel to the world-volume of  the brane-antibrane system
expands with an accelerated expansion rate while the interbrane
distance slowly decreases~\cite{Dvali:1998pa}. When the interbrane
distance becomes as short as the local string length, a tachyon
appears and starts rolling. This process of brane-antibrane
annihilation releases sufficient energy to reheat our universe and
starts the hot big-bang cosmology.

In hybrid inflation in field
theory~\cite{Linde:1991km,Linde:1993cn}, the reheating process
begins with the so called tachyonic preheating~\cite{Felder:2000hj}.
This highly non-linear, inhomogeneous dynamical process occurs due
to the tachyonic instability near the top of the water-fall
potential. The tachyonic instability converts most of the energy
into that of colliding classical waves very rapidly, within a single
oscillation. The tachyonic preheating is a typical example of
processes in which inhomogeneities are crucial.

As in the tachyonic preheating of the field-theoretic hybrid
inflation, inhomogeneities are expected to play essential roles also
in the reheating process of the warped brane inflation. In view of
this, the brane-antibrane annihilation should begin locally in regions
where the interbrane distance reaches the string length earlier than the
other parts, and those regions should turn into ``expanding holes`` on
the world-volume of the brane-antibrane system. The whole annihilation
process proceeds as those ``expanding holes'' collide with each other
and percolate. Therefore, the study of inhomogeneous annihilation of
brane-antibrane pair is an important subject towards our better
understanding of the end of the brane inflation and the beginning of the
thermal history of our universe.

Motivated by these, in this paper we would like to analyze the real
time evolution of a brane-antibrane system after the local
annihilation or recombination occurs (refer to \cite{HaNa} for an
analysis of the process of the recombination). We first present a
simple exact solution to the DBI action of D-branes which describes
the time evolution of a brane-antibrane system. Then we will
calculate the quantum energy flux produced by the open string
creation. In particular, we will mainly focus on the brane-antibrane
system of a D-string and an anti D-string (i.e. $D1-\ov{D1}$ system)
for simplicity by a technical reason that we can apply the two
dimensional conformal field theory. The argument of general
Dp-branes can also be possible in principle and will not be
essentially different. As we will show, our computation of energy
flux has a close relation to the analysis of the moving mirror
\cite{Birrell:1982ix,Ford:1997hb}. Finally, we will interestingly
notice that we encounter a negative energy flux in this process of
decaying branes.

The inhomogeneous annihilation of D-string and anti D-string
considered in this paper may be regarded as a lower-dimensional
proto-type of  the ``expanding hole'' mentioned above \cite{CaMa}.  The
process considered in this paper is probably relevant also for the
dynamics of cosmic superstring network formed after the warped brane
inflation. Indeed, the recombination of strings is one of the most
important factors for the evolution of the cosmic superstring
network. (See ref.~\cite{Polchinski:2004ia} and references therein
for a review of cosmic superstrings.)

This paper is organized as follows. In section two, we present an
exact solution which describes the decay of brane-antibrane systems.
In section three, we study the induced metric on this time-dependent
world-volume of the decaying brane. In section four, we compute the
energy flux in this process induced by the particle creation
mechanism, relating this calculation to the one in the moving mirror
system. We also discuss its string theoretic interpretations. In
section five we summarize conclusions and discuss future problems.

\section{Real Time Decay of D-string anti D-string Pair}
\setcounter{equation}{0} {\hspace{5mm}} Consider a parallel D-string
anti D-string pair separated by a distance $d$. When $d>2\pi l_s$ in
bosonic string (or $d>\s{2}\pi l_s$ in superstring), the open string
between the D-string and the anti D-string does not have any tachyonic modes
($l_s$ is the string length). Therefore in this case, the decay will
be induced by their relative motion.

Even though we started with the system of a brane and an antibrane,
we can regard the total system as a single D-string after the
recombination occurs. Then we can describe the motion of the system
by the DBI action of D-string as long as the stringy corrections are
negligible.

\subsection{Exact Solution of Time Evolution}

We assume that the D-string is included within the $(2+1)$
dimensional hypersurface $R^{1,2}$, specified by the coordinate
$(t,x,y)$ in the $D=10$ (or $D=26$) dimensional superstring (or
bosonic string) theory. The profile of the D-string is described by
the function $y=y(t,x)$. Then the Lagrangian of the DBI action for
the D-string is written as follows \be L=\s{1-\dot{y}^2+y'^2,}
\label{dbi}\ee where $\dot{y}=\f{\de y}{\de t}$ and $y'=\f{\de
y}{\de x}$.

 Let us find its solution by assuming the form
$y(t,x)=f(x)+g(t)$. The equation of motion for (\ref{dbi}) leads to
\be \ddot{g}(1+f'^2)+f''(\dot{g}^2-1)=0. \ee Thus we find
$\f{\ddot{g}}{1-\dot{g}^2}=\f{f''}{1+f'^2}=$const. This can be
solved as follows \be g(t)=\ap\log \cosh\f{t}{\ap},\ \ \ \ \
f(x)=-\ap\log \cos\f{x}{\ap}. \ee

In this way we find the time-dependent solution \be
e^{y/\ap}\cos(x/\ap)=\cosh(t/\ap),\label{scherk}  \ee where $x$ runs
the range $-\f{\pi}{2}\ap\leq x\leq \f{\pi}{2}\ap$. Actually, its
Euclidean version $e^{y/\ap}\cos(x/\ap)=\cos(z/\ap)$ is known as the
Scherk minimal surface in differential geometry.

This solution (\ref{scherk}) represents a decay of a D-string and an anti
D-string. Indeed, when $y$ is very large at a fixed time $t$, the
coordinate $x$ approaches to either of two values $x=\pm
\f{\pi}{2}\ap$. These two branches correspond to the D-string and the
anti D-string separated by the distance $d=\pi\ap$. Since we assume
that there are no tachyons between them, we require $\ap\gg l_s$.
Notice also that $y$ satisfies $y\geq \ap\log\cosh(t/\ap)$ in
(\ref{scherk}). This means that this string pair is annihilated for
$y\ll \ap\log\cosh(t/\ap)$. They are recombined at
$y=\ap\log\cosh(t/\ap)$. In summary, this D-brane at a fixed time
looks like a hairpin\footnote{In the presence of a linear dilaton,
the hairpin shaped D-brane is allowed as a static configuration
\cite{hairpin} as opposed to our case.} as described in the
Fig.\ref{fig:clip}.

\begin{figure}
\begin{center}
\includegraphics[height=4cm]{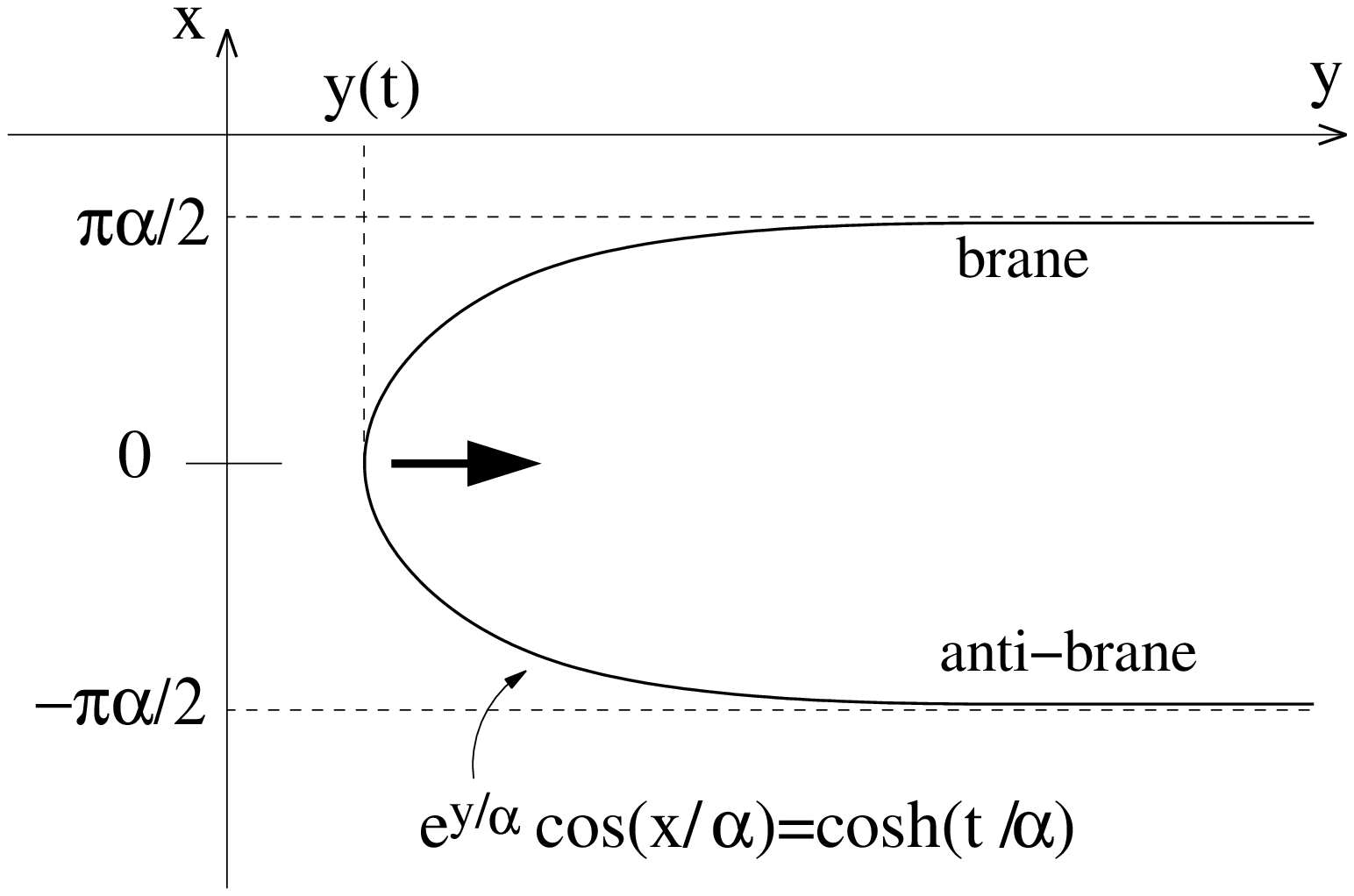}
\hspace{17mm}
\includegraphics[height=4cm]{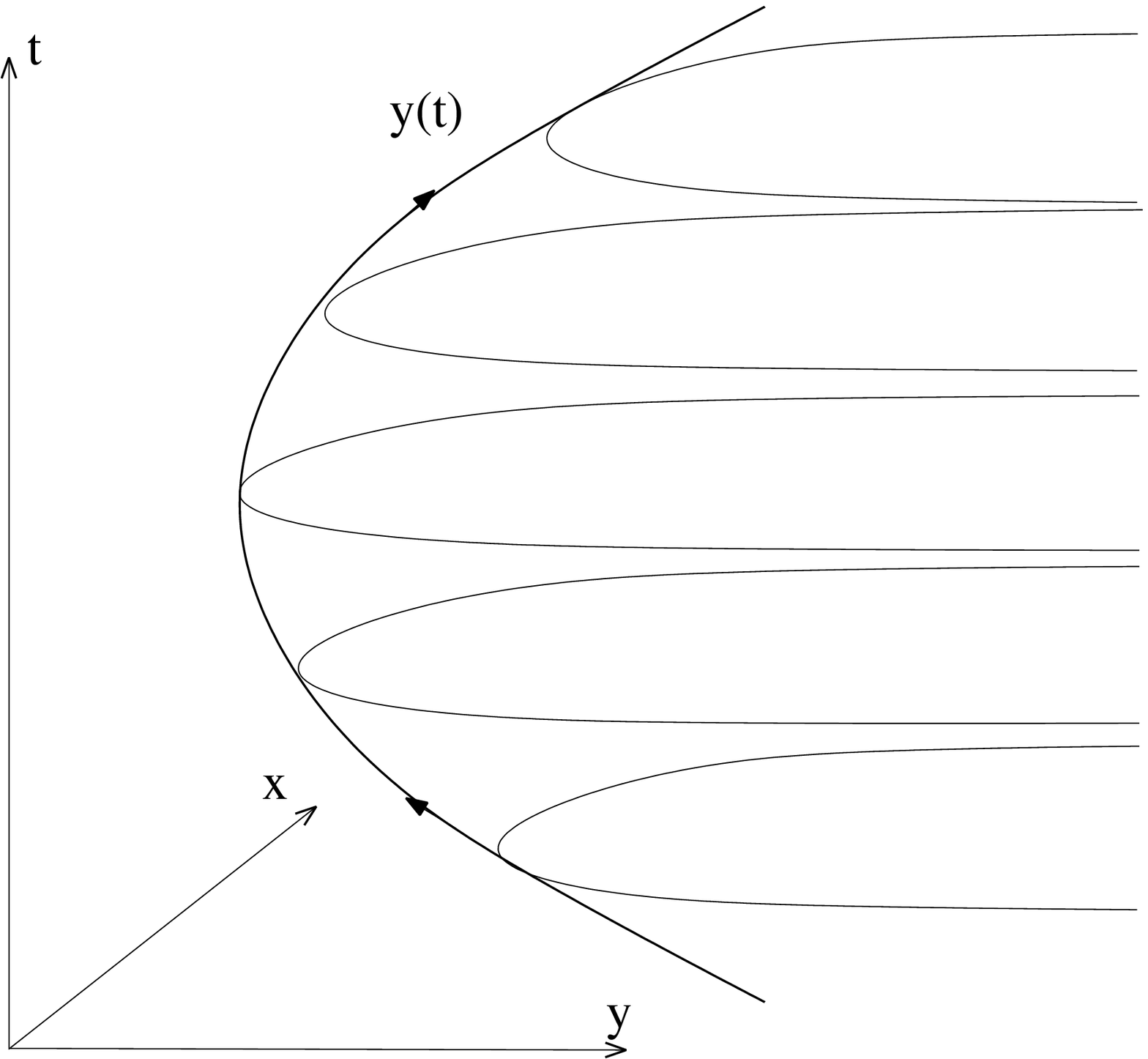}
\end{center}
\caption{The left figure shows the hairpin-like shape of the
time-dependent D-string (\ref{scherk}) at a fixed time $t$. The
right one describes its time-dependence explicitly. The tip (or
turning point) of the D-string is located at $x=0$ and its time
evolution is given by $y(t)=\ap\log \cosh(t/\ap)$. This trajectory
becomes light-like in the limit $t\to \pm\infty$. \label{fig:clip} }
\end{figure}

 The effective action of curved D-branes is described by the DBI
 action (\ref{dbi})
 only if we can neglect the higher derivative terms.
 This requires that the curvature of
the induced metric is smaller than the string scale. As we will
briefly discuss in the next section, such stringy corrections become
important only if we consider the behavior near $t\to \pm\infty$ in
the region close to the recombined point (or turning point) $x=0$.
Thus the stringy corrections are not essential to our main results
in this paper.

One may notice that another possible decay process is that the brane
and the antibrane approach via the attractive force
between them. A simple analysis estimates the life time due to this
process as $\sim \f{\ap^4}{\s{g_s}l_s^3}$. This is clearly much
longer than the time scale $\ap$ of our solution (\ref{scherk}).

We can take into account the gauge field $A_{\mu}$ on the brane. A
solution with an electric flux is shown in the in the appendix
\ref{solwithee}. We can also generalize straightforwardly the above
analysis to the systems of a Dp-brane and an anti Dp-brane by
trivially adding extra coordinates. The profile (\ref{scherk})
remains the same in this case. In these higher dimensional examples,
there are also other decay modes such as making a hole leading to a
wormhole like configuration as discussed in \cite{CaMa}. We leave
the analysis of the time evolution of such configurations for an
interesting future problem.

\section{Induced Metric and Transverse Scalars}
\setcounter{equation}{0} {\hspace{5mm} Open strings on the D-string
probe the time dependence of the background via the time-dependent
induced metric of the D-string \be ds^2=-\f{dt^2}{\cosh^2
(t/\ap)}+\f{dx^2}{\cos^2 (x/\ap)}+2\tan(x/\ap)\tanh(t/\ap)dtdx.
\label{indm} \ee

\subsection{Induced Metric in Conformal Gauge}

It is useful to rewrite this metric into the one in the conformal
gauge \be ds^2=A(\tau,y)(-d \tau^2+dy^2), \label{ind} \ee so that
the Lagrangian of minimally coupled massless scalar fields $\phi$
becomes simple at quadratic order \be
L=\f{1}{2}\left(-(\de_{\tau}\phi)^2+(\de_{y}\phi)^2)\right).\label{masslessf}
\ee Requiring that the coordinate $y$ in (\ref{ind}) is the same as the one
in (\ref{scherk}), we find an appropriate coordinate transformation
as follows \ba \tau&=&\ap\log
\left[\s{1+\sinh^2(t/\ap)\sin^2(x/\ap)}+\sin(x/\ap)\sinh(t/\ap)\right],
\no y&=&\ap\log \cosh(t/\ap)-\ap \log \cos(x/\ap). \ea It is
equivalent to the relation \be
\sinh(\tau/\ap)=\sinh(t/\ap)\sin(x/\ap),\ \ \ \ \ \ \ \
e^{y/\ap}=\f{\cosh(t/\ap)}{\cos(x/\ap)}. \ee The metric in this
coordinate system reads \be
ds^2=-\left(\f{1}{2}+\f{\cosh\f{v-u}{2\ap}}
{2\s{\sinh\f{v}{\ap}\cdot\sinh\f{-u}{\ap}}}\right)dudv
,\label{induced} \ee where we defined \be u=\tau-y,\ \ \ \ v=\tau+y.
\ee

We need to deal with the metric (\ref{induced}) carefully. It is
well-defined for $y>|\tau|$. When we go beyond this bound, the
coordinate $y$ (and $\tau$) in (\ref{induced}) becomes time-like
(space-like) and we have to exchange $\tau$ with $y$ and vise versa,
which is equal to the sign flip of $u$ and $v$. Also at the bound
$y=\pm \tau$, the metric becomes singular. Actually, we can further
perform a coordinate transformation and obtain a completely smooth
metric which covers the total spacetime on the D-string as we show
in the appendix \ref{globalm}.

\subsection{Asymptotically Flat Coordinate}

The metric (\ref{induced}) at the null past infinity $u=-\infty$
approaches \be
A(u,\infty)=\f{1}{2}\left(1+\f{1}{\s{1-e^{-2v/\ap}}}\right). \ee On
the other hand in the limit $v=\infty$ of the future null infinity
we get \be
A(-\infty,v)=\f{1}{2}\left(1+\f{1}{\s{1-e^{2u/\ap}}}\right). \ee
Thus we can define the asymptotically flat metric by the coordinate
transformation \ba &&V=-\f{\ap}{2}\log (1-\s{1-e^{-2v/\ap}}),\no &&
U=\f{\ap}{2}\log (1-\s{1-e^{2u/\ap}})\label{flcor}. \ea In terms of
$(U,V)$, the metric $ds^2=-A(U,V)dUdV$ is found to be
\begin{eqnarray}
 \ti{A}(U,V)
  & = & A(u,v)\frac{dv}{dV}\frac{du}{dU} \label{flatm} \\
  & = &
  \frac{2\left(1+e^{-V/\alpha}e^{U/\alpha}\sqrt{2-e^{-2V/\alpha}}\sqrt{2-e^{2U/\alpha}}
  + (1-e^{-2V/\alpha})(1-e^{2U/\alpha})\right)}
  {(2-e^{-2V/\alpha})(2-e^{2U/\alpha})}\nonumber .
\end{eqnarray}
It is easy to see that this is manifestly asymptotically flat:
%
\begin{equation}
 ds^2 \simeq -dUdV
\end{equation}
near the past null infinity ($U\to -\infty$) and the future null
infinity ($V\to\infty$). This coordinate system is regular in the
regime
%
\begin{equation}
 V > -\frac{\alpha}{2}\ln 2, \quad U < \frac{\alpha}{2}\ln 2.
  \label{eqn:regular-region}
\end{equation}

\subsection{Global Geometry}

 In this coordinate $(U,V)$, the turning point $x=0$ is described by the
curve \be e^{\f{-2V}{\ap}}+e^{\f{2U}{\ap}}=2,\ee or \begin{equation}
 V = P(U), \quad P(U) \equiv
  -\frac{\alpha}{2}\ln\left(2-e^{2U/\alpha}\right).
  \label{eqn:P(U)}
\end{equation}
The half of the D-string spacetime $x\geq 0$ is mapped to the region
$V\geq P(U)$, while another half $x\leq 0$ is mapped to the region
$U\geq P(V)$. Since they are actually equivalent by the reflection,
the global geometry is obtained by gluing two copies of the region
$V>P(U)$ along the symmetry axis $V=P(U)$ as in
Fig.\ref{fig:penrose}. Note that both of them are included in the
regime (\ref{eqn:regular-region}) of validity of the coordinates
$(U,V)$. For more explanations of the global structure of our
metric, see the appendix \ref{globalm}.

\begin{figure}
\begin{center}
\includegraphics[height=7cm]{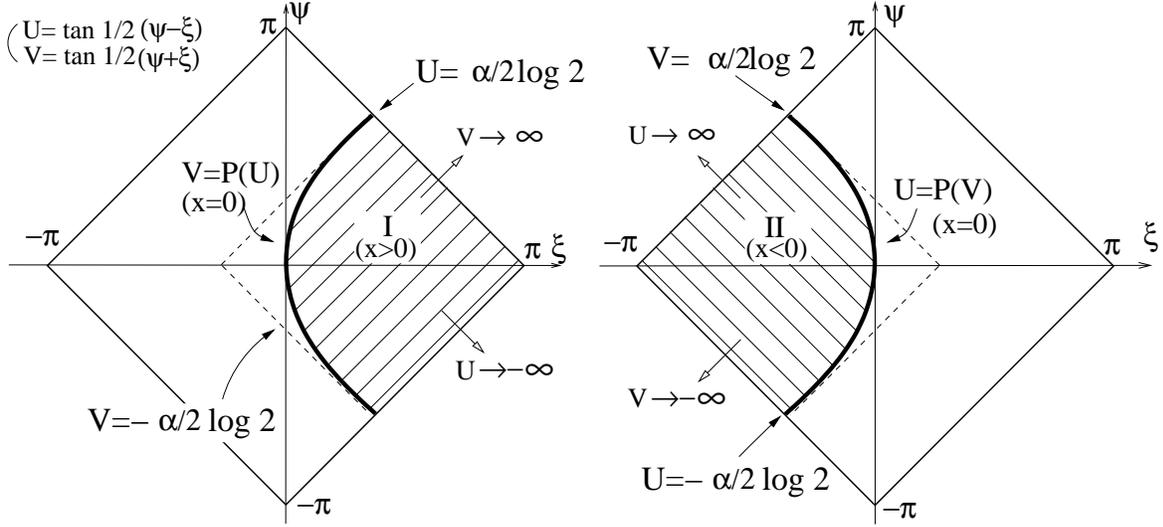}
\end{center}
\caption{The Penrose diagram of our time-dependent spacetime
 (\ref{flatm}). $x\ge0$ is mapped to region I, and  $x \le 0$ is mapped
 to region II. The global geometry can be obtained by gluing two shaded
 regions along the curve $x=0$ represented by the thick curve.
 \label{fig:penrose} }
\end{figure}

If we define $U=\bar{t}-\bar{y}$ and $V=\bar{t}+\bar{y}$, this curve
is described by \be \bar{y}=\f{\ap}{2}\log\cosh\f{2\bar{t}}{\ap}.
\label{curve} \ee Notice that this curve is slightly off from the
naive estimate $y\sim \ap\log\cosh\f{\bar{t}}{\ap}$ from
(\ref{scherk}) by the factor 2. This is related to the fact that the
value of the metric function $\tilde{A}$ on $V=P(U)$ is off from unity,
%
\begin{equation}
 \tilde{A}(U,P(U)) = 4.
\end{equation}

 On the other hand, along the symmetry axis
$dU$ and $dV$ are related as
%
\begin{equation}
 dU = \frac{e^{-2V/\alpha}}{2-e^{-2V/\alpha}}dV.
\end{equation}
Thus, the induced line element along the symmetry axis is
%
\begin{equation}
 ds^2 =  -\frac{4e^{-2V/\alpha}}{2-e^{-2V/\alpha}}dV^2.
\end{equation}
This shows that the symmetry axis $V=P(U)$ is asymptotically null in
the limit $V/\alpha\to\infty$.

One can check that the extrinsic curvature of the symmetry axis
vanishes so that the two copies of the region $V>P(U)$ are glued
smoothly.  For this purpose, let us introduce a unit normal vector
$n^a$ as
%
\begin{equation}
 n^U = -\frac{1}{\sqrt{\tilde{A}}}e^{-U/\alpha}e^{-V/\alpha}, \quad
 n^V = \frac{1}{\sqrt{\tilde{A}}}e^{U/\alpha}e^{V/\alpha}.
\end{equation}
The extrinsic curvature is defined as
%
\begin{equation}
 K_{ab} = \left.\frac{1}{2}q_a^{\ c}q_b^{\ d}
  (\nabla_c n_d+\nabla_d n_c)\right|_{V=P(U)},
\end{equation}
where
%
\begin{equation}
 q_{ab} = g_{ab} - n_an_b.
\end{equation}
By explicit calculation it is shown that
%
\begin{equation}
 K_{ab} = 0.
\end{equation}
Therefore, two copies of the region $V>P(U)$ are glued smoothly as
should be. Hereafter, we call one of the two copies of the region
$V> P(U)$ the region I and the other the region II (see Fig.2.). In
this way we get a completely smooth coordinate which covers the
total spacetime.

Before we proceed, we would also like to examine stringy corrections
which we neglected in this paper. In order to argue it is
negligible, we need to require $l^2_s R_{\mu\nu\rho\sigma}^{(2)}\ll 1$
in the region I (the analysis of region II is the same). The
curvature tensor takes larger values as we come close to the curve
$V=P(U)$ and there it behaves like
$R_{\mu\nu\rho\sigma}^{(2)}\sim \f{1}{\ap^2}e^{4V/\ap}$. Therefore the
condition of neglecting stringy corrections is given by
\be V\ll \f{\ap}{2}\log\f{\ap}{l_s}.\label{stringy} \ee
Since we assumed from the beginning $\ap\gg l_s$, we can still neglect
the stringy corrections even if we take $\f{V}{\ap}$ to be appropriately
large.

\subsection{Transverse Scalar as Tachyon}

The main motivation of the above analysis of the induced metric is
to understand the low energy effective action of open string modes.
Since gauge fields in $1+1$ dimension are not dynamical, we would
like to consider the transverse scalar fields. First we focus on the
transverse scalars $\phi_i$ $(i=3,4,\ddd ,D)$ which correspond to
the deformations into the directions orthogonal to the $R^{1,2}$
defined by $(t,x,y)$. In the low energy limit, their effective
actions are all equal to the one of the (minimally coupled) massless
scalar field $L=\f{1}{2}\s{-g}g^{\mu\nu}\de_{\mu}\phi\de_{\nu}\phi$,
where $\mu$ and $\nu$ run over the D-string world-volume and
$g_{\mu\nu}$ is the induced metric of the D-string (\ref{flatm}).

There is another transverse scalar which describes the motion of
D-string within the $R^{1,2}$ spacetime. As shown in the appendix
\ref{DBIa}, at the quadratic order, the effective action is given by
the one of a massive scalar field
$L=\f{1}{2}\s{-g}(g^{ab}\de_{a}\phi\de_{b}\phi+m^2\phi^2)$ with the
mass square given by \be m^2=\f{1}{2}(R^{(2)}-K^{\mu\nu}K_{\mu\nu}), \ee
where $K_{\mu\nu}$ is the extrinsic curvature.  It turns out that in
our example of the decaying D-string and anti D-string the mass
square becomes negative \be m^2=-\f{2\sec^2\f{x}{\ap}~
\mbox{sech}^2\f{t}{\ap} }{\ap^2(\sec^2\f{x}{\ap}~ \mbox{sech}^2
\f{t}{\ap}+\tan^2 \f{x}{\ap}~ \tanh^2 \f{t}{\ap})^2}<0,
\label{tachyon} \ee and therefore the scalar is a tachyonic mode. In
this paper we restrict to the case where the condensation of this
tachyon does not occur. It is a quite interesting future problem to
see what will happen if the tachyon condenses.

\section{Energy Flux from Decaying D-brane}
\setcounter{equation}{0} {\hspace{5mm}} Now we move on to the main
part of this paper. We would like to analyze the particle creation
(open string creation) in the time-dependent open string theory on
the D-string (\ref{scherk}). As we will mention, this analysis is
essentially the same as that of moving mirror systems. First we
concentrate on massless transverse scalars\footnote{We will not
discuss the tachyonic scalar (\ref{tachyon}) because its mass is
time-dependent and the quantum analysis will be much more
complicated. } and later we will discuss the result when we take all
stringy modes into account.
\subsection{Energy Flux in Massless Scalar Theory}

We introduce a minimally-coupled, massless scalar field $\phi$. By
decomposing the field into left- and right-moving parts as
%
\begin{equation}
 \phi = \left\{\begin{array}{ll}
     \phi_{I+}(V) + \phi_{I-}(U) & (\mbox{region I} )\\
     \phi_{II+}(V) + \phi_{II-}(U) & (\mbox{region II} )\\
           \end{array}
           \right. ,
\end{equation}
the smoothness of the field implies that
%
\begin{equation}
 \phi_{I-}(U) = \phi_{II+}(P(U)), \quad
  \phi_{II-}(U) = \phi_{I+}(P(U)).
\end{equation}
Thus, we obtain a general solution
%
\begin{equation}
 \phi = \left\{\begin{array}{ll}
     \phi_{I+}(V) + \phi_{II+}(P(U)) & (\mbox{region I} )\\
     \phi_{II+}(V) + \phi_{I+}(P(U)) & (\mbox{region II} )\\
           \end{array}
           \right. .
\end{equation}

We consider the vacuum $|in\rangle$ defined by the following mode
functions
%
\begin{eqnarray}
 \phi^{(L)} =
  \left\{\begin{array}{ll}
   \frac{1}{\sqrt{4\pi\omega}}e^{-i\omega V}
    & (\mbox{region I} )\\
      \frac{1}{\sqrt{4\pi\omega}}e^{-i\omega P(U)}
       & (\mbox{region II} )\\
     \end{array}
  \right. , \nonumber\\
 \phi^{(R)} =
  \left\{\begin{array}{ll}
   \frac{1}{\sqrt{4\pi\omega}}e^{-i\omega P(U)}
    & (\mbox{region I} )\\
      \frac{1}{\sqrt{4\pi\omega}}e^{-i\omega V}
       & (\mbox{region II} )\\
     \end{array}
  \right. .
\end{eqnarray}

The Weightman function in each region is
%
\begin{eqnarray}
 G(x,x') & \equiv &
  \langle in|\phi(x)\phi(x')|in\rangle \nonumber\\
 & = &
  \int_0^{\infty}d\omega
  \left[\phi^{(L)}_{i\omega}(x)\phi^{(L)*}_{i\omega}(x') +
   \phi^{(R)}_{i\omega}(x)\phi^{(R)*}_{i\omega}(x')\right]
  \nonumber\\
  & = & G^{(0)}(x,x') + \Delta G(x,x') ,
\end{eqnarray}
where
%
\begin{equation}
 G^{(0)}(x,x') \equiv
  \int_0^{\infty}\frac{d\omega}{4\pi\omega}
  \left[e^{-i\omega(V-V')}+e^{-i\omega(U-U')}\right] ,
\end{equation}
and
%
\begin{eqnarray}
 \Delta G(x,x') & \equiv &
  G(x,x') - G^{(0)}(x,x') \nonumber\\
  & = & \int_0^{\infty}\frac{d\omega}{4\pi\omega}
  \left( e^{-i\omega(P(U)-P(U'))}-e^{-i\omega(U-U')}\right)
  \nonumber\\
  & = & \frac{1}{4\pi}\ln
   \left[\frac{U-U'+i\epsilon}{P(U)-P(U')+i\epsilon}\right].
\end{eqnarray}
Here, to obtain the last line, we have used
%
\begin{equation}
 \int_0^{\infty}\frac{d\omega}{\omega}
  \left(e^{i\omega x}-e^{i\omega y}\right) =
  \ln\left(\frac{y-i\epsilon}{x-i\epsilon}\right).
\end{equation}

Thus, the renormalized stress-energy tensor for each field for the
state $|in\rangle$ is decomposed as
%
\begin{equation}
 T^{(ren)}_{ab} = T^{(0)}_{ab} + \Delta T_{ab},
\end{equation}
where $T^{(0)}_{ab}$ is the contribution from $G^{(0)}$ obtained
later and $\Delta T_{ab}$ is the contribution from $\Delta G$ given
by
%
\begin{eqnarray}
 \Delta T_{UU} & = & \lim_{x'\to x}\partial_U\partial_{U'}
  \Delta G(x,x')
  = -\frac{1}{24\pi}\left\{ P(U),U\right\},\nonumber\\
 \Delta T_{VV} & = & \lim_{x'\to x}\partial_V\partial_{V'}
  \Delta G(x,x') = 0, \nonumber\\
 \Delta T_{UV} & = & \Delta T_{VU} = 0.
\end{eqnarray}
where $\{f,z\}$ is the Schwarzian derivative defined by
%
\begin{equation}
 \{f,z\} = -2\sqrt{df/dz}\frac{d^2}{dz^2}
  \left(\frac{1}{\sqrt{df/dz}}\right),
\end{equation}
and satisfies
%
\begin{equation}
 \left(\frac{f(z_2)-f(z_1)}{x_2-x_1}\right)^2
  \frac{1}{f'(z_1)f'(z_2)}
  = 1 - \frac{1}{6}\left\{f,z\right\}(z_2-z_1)^2
  + O(z_2-z_1)^4.
\end{equation}

The contribution $T^{(0)}_{ab}$ from $G^{(0)}(x,x')$ can be obtained
by properly subtracting divergent parts, differentiating with
respect to $x$ and $x'$ and taking the coincident limit. The result
must be local in the sense that all components are functions of the
metric function $\tilde{A}$ and its derivatives. In particular, they
must be independent of the function $P(U)$.  Moreover,
$T^{(0)}_{ab}$ must vanish in the asymptotically flat region since
the subtracted divergent parts are local geometric functions.

Actually, all components of $T^{(0)}_{ab}$ can be determined by the
following three conditions: (i) it satisfies the conservation
equation $\nabla^aT^{(0)}_{ab}=0$; (ii) its trace is determined by
the trace anomaly $T^{(0)a}_{\ a}=R/(24\pi)$, where
$R=4\tilde{A}^{-3}[\tilde{A}\partial_U\partial_V\tilde{A}-(\partial_U\tilde{A})(\partial_V\tilde{A})]$
is the Ricci scalar; (iii) it vanishes in the  asymptotically flat
region. The general solution to the conditions (i) and (ii) is
%
\begin{eqnarray}
 T^{(0)}_{UU} & = & -\frac{1}{12\pi}\sqrt{\tilde{A}}\partial_U^2
  \left(\frac{1}{\sqrt{\tilde{A}}}\right) + \alpha(U), \nonumber\\
 T^{(0)}_{VV} & = & -\frac{1}{12\pi}\sqrt{\tilde{A}}\partial_V^2
  \left(\frac{1}{\sqrt{\tilde{A}}}\right) + \beta(V), \nonumber\\
 T^{(0)}_{UV} & = &  T^{(0)}_{VU}  = -\frac{\tilde{A}}{96\pi}R,
\end{eqnarray}
where $\alpha(U)$ and $\beta(V)$ are arbitrary functions. Finally,
the condition (iii) implies that
%
\begin{equation}
 \alpha(U) = 0, \quad \beta(V) = 0.
\end{equation}

In summary, we obtain
%
\begin{eqnarray}
 T^{(ren)}_{UU} & = & -\frac{1}{12\pi}\sqrt{\tilde{A}}\partial_U^2
  \left(\frac{1}{\sqrt{\tilde{A}}}\right) + \Delta T_{UU}, \nonumber\\
 T^{(ren)}_{VV} & = & -\frac{1}{12\pi}\sqrt{\tilde{A}}\partial_V^2
  \left(\frac{1}{\sqrt{\tilde{A}}}\right) , \nonumber\\
 T^{(ren)}_{UV} & = &  T^{(0)}_{VU}
  = \frac{1}{24\pi\tilde{A}^2}
  \left[(\partial_U\tilde{A})(\partial_V\tilde{A})
   -\tilde{A}\partial_U\partial_V\tilde{A}\right],
\end{eqnarray}
where
%
\begin{equation}
 \Delta T_{UU} =
  -\frac{1}{24\pi}\left\{ P(U),U\right\}.
\end{equation}
In the asymptotically flat region, only $\Delta T_{UU}$ is relevant:
%
\begin{eqnarray}
 T^{(ren)}_{UU} & \to & \Delta T_{UU}, \nonumber\\
 T^{(ren)}_{VV} & \to & 0, \nonumber\\
 T^{(ren)}_{UV} & = &  T^{(0)}_{VU}  \to 0.
\end{eqnarray}

For $P(U)$ in (\ref{eqn:P(U)}), the Schwarzian derivative is
%
\begin{equation}
 \left\{ P(U),U\right\} = -\frac{8}{\alpha^2}
  \frac{1-e^{2U/\alpha}}{(2-e^{2U/\alpha})^2}.
\end{equation}
Thus, we obtain
%
\begin{equation}
 \Delta T_{UU} = \frac{1}{3\pi\alpha^2}
  \frac{1-e^{2U/\alpha}}{(2-e^{2U/\alpha})^2}.\label{fluxt}
\end{equation}

Notice that this energy flux (\ref{fluxt}) approaches the following
constant in the limit $U\to -\infty$ \be \Delta T_{UU}\to
\f{1}{12\pi\ap^2}\equiv\f{\pi}{12}T_{D}^2.\label{efm}\ee This
suggests that a static detector will observe the thermal flux at the
temperature $T_{D}=\f{1}{\pi\ap}$ long before it will collides with
the tip of the D-string (i.e. $x=0$). This can be more clearly
confirmed by computing the Bogolubov coefficients as we do in the
appendix \ref{apbo}.

However, later it gradually decreases and eventually changes its
sign (i.e. negative flux) as described in Fig.\ref{fig:timeflux}. In
the end the detector collides with the tip of D-string. Thus the
negative flux is observed only for a finite time $\Delta
t=\f{\ap}{2}\log 2$.

It even gets negatively divergent precisely when it collides with
the tip of the D-string at $U=\f{\ap}{2}\log 2$ if we literally
accept (\ref{fluxt}). However, we have to remember that the string
corrections become important when the condition (\ref{stringy}) is
violated. Indeed, if we assume $V\sim \f{\ap}{2}\log \f{\ap}{l_s}$,
we find that the energy density becomes of the same order as the
fundamental string tension $\Delta T_{UU}\sim \f{1}{l^2_s}$.
Therefore when this divergence happens the stringy corrections
cannot already be negligible and they will plausibly give
substantial modifications.

\begin{figure}
\begin{center}
\includegraphics[height=6cm]{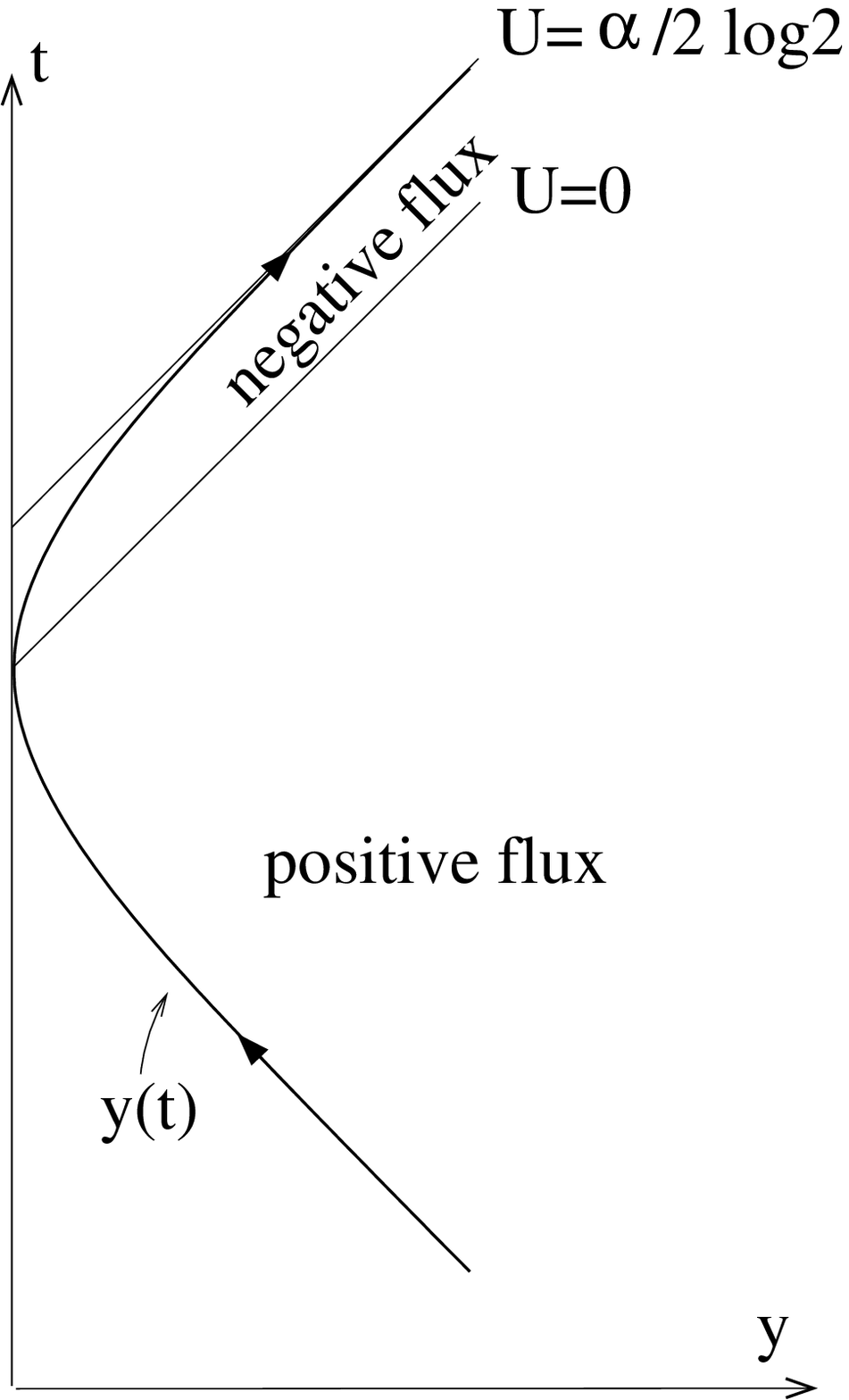}
\hspace{10mm}
\includegraphics[height=6cm]{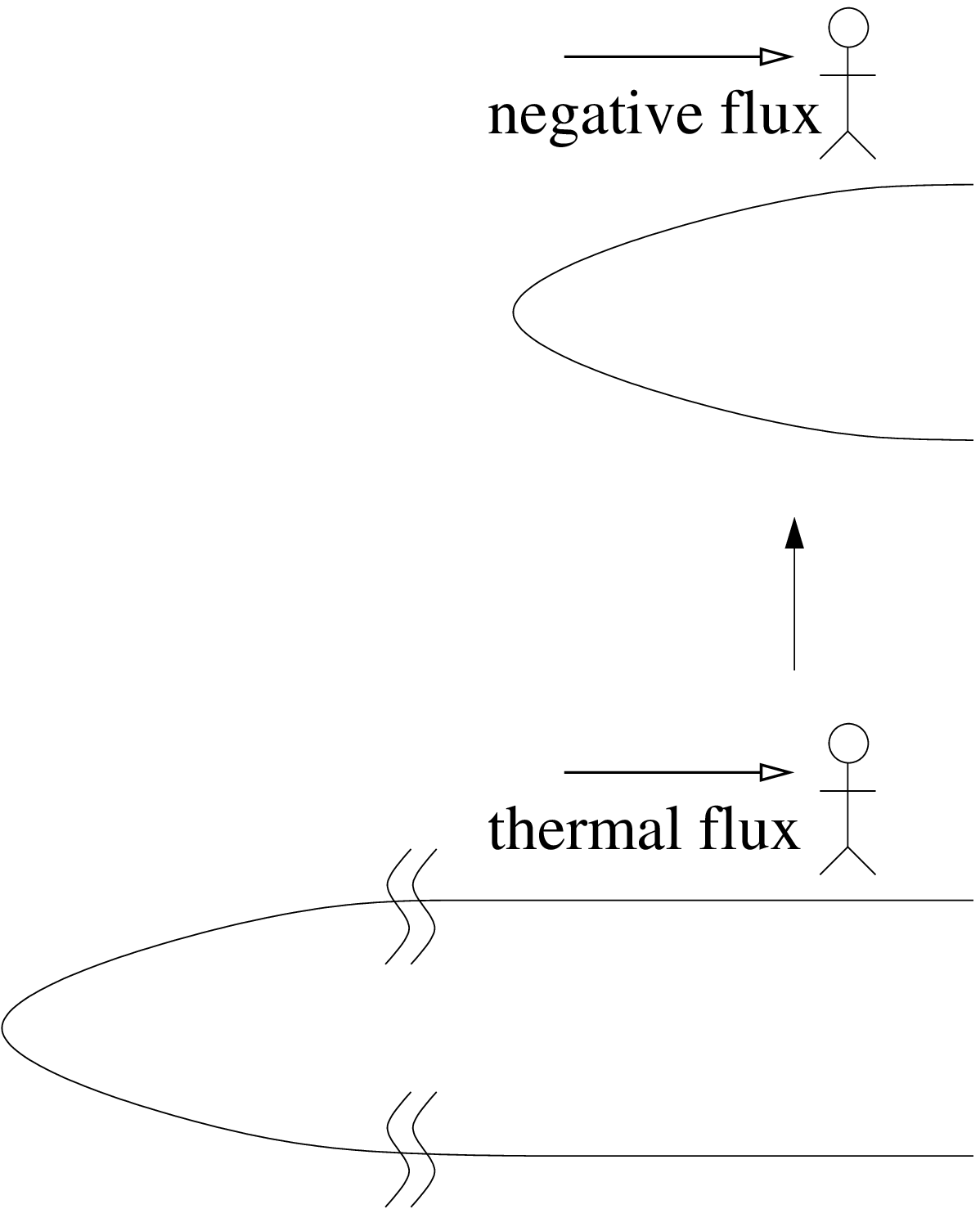}
\end{center}
\caption{The behavior of the energy flux due to the particle
creation under time evolution. In the left figure we show the
spacetime regions where the sign of the energy flux is $+$ or $-$. A
detector which is far from the tip of the brane first observes a
thermal energy flux. However, later it changes into a negative
energy flux and finally collides with the end point. This is
summarized in the right figure. \label{fig:timeflux} }
\end{figure}

\subsection{D-brane Interpretation and Moving Mirror}

Now we would like to interpret the above result from the viewpoint
of the decay of the D-string-anti D-string system. Before the decay
occurs, the effective world volume theory of the system at low energy is regarded as a $1+1$ dimensional
$U(1)\times U(1)$ gauge theory. Concentrating on a massless scalar
in this theory, the ones for the brane and antibrane are denoted by
$\phi_{I}(t,y)$ and $\phi_{II}(t,y)$. Here notice that we considered
that the fields on the D-brane and the anti D-brane live in the same
$1+1$ dimensional spacetime assuming that the distance between the
D-branes is much smaller\footnote{We still assume that the distance
is much larger than string scale.} than the typical length scale
which the observer probes.

After the time evolution (\ref{scherk}) starts, the brane and
anti-brane become connected at a point $y=y_{*}(t)$ and the
$U(1)\times U(1)$ gauge symmetry is broken there. The continuity of
the scalar fields at $y=y_{*}(t)$ requires the following boundary
conditions  \be \phi_{I}(t,y_*(t))-\phi_{II}(t,y_*(t))=0,\ \ \ \
\left.n^a\de_{a}\left[\phi_{I}(t,y)+\phi_{II}(t,y)\right]\right|_{y=y_*(t)}=0,
\label{boundaryc} \ee where $n^a$ is a unit normal to the curve
$y=y_*(t)$. Taking the global metric, the function $y_*(t)$ in our
case is explicitly given by (\ref{curve}).

In our global geometry, the massless particle which propagates in
the $x<0$ (or $x>0$) region "reflects" at $y=y_{*}(t)$, and then
comes into the $x>0$ (or $x<0$) region. Thus we can consider the
trajectory $y=y_{x}(t)$ as a perfect "mirror" in the $1+1$
dimensional space-time, and this setup is essentially the same as
the one of the moving mirror system in which boundary conditions
depend on time (see e.g.\cite{Birrell:1982ix, Ford:1997hb}).

In general situations, when the mirror is moving at a non-zero
acceleration, the static observer detects the thermal flux by the
Unruh effect. It is described by a quantum field theory in the
presence of the time-dependent boundary condition. For example, the
Hawking radiation from a Schwarzschild black hole can also be
regarded as a particular setup of moving mirror defined by
$y(t)=-t-e^{-2\pi T_{bh}t}$, where $T_{bh}$ is the Hawking
temperature of the black hole. A similar analysis of the energy flux
$\Delta T_{UU}$ in section 3.1 and of the Bogolubov coefficients in
the appendix \ref{apbo}, correctly reproduces the well-known
properties of the Hawking radiation \cite{Birrell:1982ix}.

Indeed, our boundary conditions (\ref{boundaryc}) are equivalent to
the moving mirror system with two massless scalars. One of the two
scalars obeys the Dirichlet boundary condition and the other does
the Neumann one. Therefore, the energy flux in the low energy world
volume theory of the brane-antibrane system should be identified
with the twice of (\ref{fluxt}). If we consider other massless
fields, assuming the conformal invariance, we just need to multiply
the value of central charge $c$ with (\ref{efm}).

In the actual open string theory, there are infinitely many massive
fields in the brane-antibrane system in addition to the massless
fields we discussed. Let us try to take all stringy massive modes
into account in our argument of particle creation. Notice also that
if we are interested in higher dimensional Dp-branes ($p>1$) case,
we can regard the transverse momenta as the Kaluza-Klein momenta,
generating extra mass terms. Therefore the treatment of decaying
Dp-branes is similar. Unfortunately, even in two dimension, an exact
analysis of a massive scalar field in the moving mirror setup is not
available at present. Still we can roughly estimate the total energy
flux as follows.

We consider the open string creation long before the observer
collides with the tip of the brane. The result (\ref{fluxt}) argues
that we will get a thermal radiation at the temperature
$T_{D}=\f{1}{\pi\ap}$. We can show this explicitly by computing the
Bogolubov coefficients from the reflection condition (\ref{curve})
as in the appendix \ref{apbo} of this paper. The string production
rate is roughly estimated by multiplying the Hagedron density of
states ($T_H$ is the Hagedron temperature) \be \sim\int dE ~
e^{\f{E}{T_{H}}}e^{-\f{E}{T_D}}. \ee Notice that open string modes
become in general massive, but their masses clearly suppress the
particle creation and so the above calculation gives the upper
bound. Thus we can conclude that the total string creation is (UV)
finite since $T_{H}\sim \f{1}{l_s}\gg T_{D}$ under the assumption.
This result makes a sharp contrast with the open and closed string
creation in the rolling tachyon backgrounds
\cite{St,LaLiMa,StTa,Sen,hairpin}, where the string creation rate
diverges.

\subsection{Analogue of Unruh Vacuum and Negative Energy Flux}

In the actual decay processes of brane-antibrane systems, the time
evolution like (\ref{scherk}) will be triggered by tunneling
processes or by collisions with other objects such as closed strings
or D-branes. From this viewpoint, it is more natural to assume that
when $t<0$ the D-brane remains static and that the real time
evolution (\ref{scherk}) starts at the time $t=0$. We may think that
this is analogous to the Unruh vacuum in an analogy to the Hawking
radiation from black holes. We can equally treat this case using the
conformal mapping technique. The upshot is that the energy flux for
$U<0$ becomes vanishing and that there remains the negative flux
when $U>0$ as in Fig.\ref{fig:enetwo}. This can easily be understood
from the causal evolution. If we want to explicitly confirm this, it
is useful to start with a smooth time-dependent evolution which is
very similar to this. For example, we can choose the trajectory of
the turning point as \be \f{y}{\alpha} = \frac{1}{2} \log
(e^{\frac{2t}{\alpha}}+1).\ee Notice that now our new $y$ converges
for $t\ll 0$ as $y \sim 0$. We eventually find the following energy
flux
\begin{equation}
T_{uu}= \frac{1}{12\pi \alpha^2} \frac{1-2\cosh
[2u/\alpha]}{\sinh(2u/\alpha)^2}<0.
\end{equation}
Notice that this flux is always negative and $T_{uu}$ diverges as
$u\to 0$ as expected\footnote{$U$ is related to $u$ via
$u=U-\f{\ap}{2}\log2.$}. A similar negative energy flux was found in
the analysis of $c=1$ matrix model dual to two dimensional string
theory \cite{KaSt}.

The negative energy flux may seem counterintuitive at first. In
fact, this phenomenon is known to be typical in the moving mirror
systems and a number of works have been done to check that it does
not violate fundamental laws in thermal dynamics and so on
\cite{Birrell:1982ix,Ford:1997hb}. A more familiar example of
negative flux will be the evaporation of black holes, which can also
be explained as the moving mirror effect.

In our setup, observers situated far from the tip will lose their
energy due to the negative energy flux for a while before they
collide with the tip. Since the total energy should be conserved,
this energy lost by the observers will be gained by the moving
D-string.

\begin{figure}
\begin{center}
\includegraphics[height=6cm]{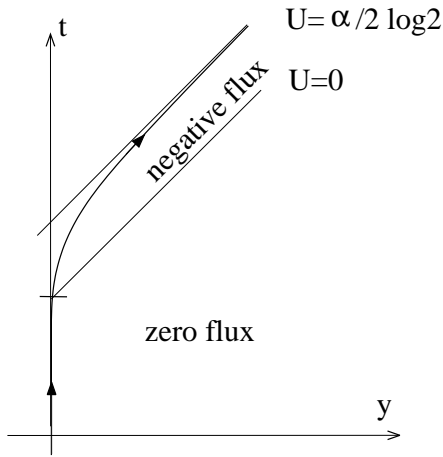}
\end{center}
\caption{The behavior of the energy flux in the vacuum analogous to
the Unruh vacuum under time evolution. Notice that the energy flux
is always non-positive.
 \label{fig:enetwo} }
\end{figure}

\vspace{10mm}

\section{Conclusions and Discussions}
\setcounter{equation}{0} {\hspace{5mm}} In this paper we analyzed an
explicit decaying process of the brane-antibrane system described by
the simple solution (\ref{scherk}) to the DBI action. At a fixed
time, its world volume takes a hairpin shape (Fig.\ref{fig:clip}).
The position of the tip (or the turning point) in this hairpin is
time-dependent. This will be one of the simplest time-dependent
backgrounds in string theory.

After we calculated the induced metric of its world volume, we
computed the quantum energy flux due to the particle creation in
this time-dependent system. We found that it looks like a thermal
flux when an observer is far from the turning point. At a later
time, slightly before the observer collides with the turning point,
the energy flux becomes negative. This means that observers lose
their energy before they collide with the tip of the D-brane. It is
an interesting future problem to investigate the physical
consequences of this phenomenon. Since the energy flux computed in
this paper comes from the open string excitations, it will be
intriguing if we can also compute closed string radiations from the
decaying brane system.

Even though the solution itself (\ref{scherk}) describe the decay of
a $Dp-\bar{D}p$ for any $p$, we mainly restrict the analysis of
energy flux to the simplest case $p=1$ for simplicity. The study of
higher dimensional case $p>1$ will also be an interesting future
problem even from the viewpoint of phenomenological applications.

In our analysis, we always described the dynamics of D-branes by the
DBI action. Therefore we neglected the stringy effects. Even though
these corrections are not important for the regime we are mostly
interested in, they become relevant in order to know what will
happen just before the observer collides with the tip. Therefore, it
would be very intriguing to construct corresponding boundary states
which are $\al$ exact.

When we calculated quantum energy flux from inhomogeneous decay of a
brane-antibrane pair, we treated the solution (\ref{scherk})  as a
fixed background and totally neglected backreaction due to the flux.
Once the backreaction is taken into account, one should then be able to
answer various physical questions. One of such questions would be
about consequences of the late-time negative energy flux.  This is,
in a sense, similar to questioning physical consequences of quantum
energy flux from a black hole. It is well known that quantum fields in a
fixed black hole background exhibit thermal flux called Hawking
radiation. Taking the backreaction of this positive energy flux into
account, it is concluded that the black hole should lose its energy,
i.e. mass, and that the flux from the hole should increase since the
hole's temperature is inversely proportional to its mass. This
catastrophic loss of the black hole mass should continue until the
system exits the regime of validity of the low energy effective field
theory.

In the case of the decaying brane-antibrane system, the late-time
energy flux is negative and this should correspond to increase of
background energy. Therefore, it is naturally expected that the motion
of the decaying brane-antibrane pair should receive extra acceleration
as a result of the negative energy flux. (Note that, even without
including the backreaction effect, the system receives acceleration
since it loses the rest mass of the decaying segment. The backreaction
due to the negative energy flux adds further acceleration to the motion
of the system.)  In particular, if we consider corrections to the
behavior shown in the right panel of Fig.\ref{fig:clip}, the tip of the
bent D-string should be further accelerated towards the $+y$ direction
at late time. It should be an important and interesting future work to
investigate the backreaction of the negative energy flux more
quantitatively.

Finally, from the string theory viewpoint,  it is very likely that
the beginning of our universe includes many D-branes and anti
D-branes. This makes the study of their annihilation processes very
important to understand the cosmology of our universe. Therefore it
will also be a quite interesting future project to consider
applications of our setups and results to cosmological scenarios.
One problem in this direction will be to analyze a similar
time-dependent D-branes in AdS or warped backgrounds.
Indeed, as stated in the introduction, the end of warped brane inflation
is annihilation of a brane-antibrane pair in a warped geometry, and
inhomogeneities of the type considered in this paper are expected to be
crucial in the (p)reheating process after inflation as well as in the
evolution of cosmic string network. The inhomogeneous and time-dependent
background presented in this paper and its extensions to higher
dimensions and/or to warped backgrounds are, therefore, of cosmological
interest in the framework of warped brane inflation. Further studies in
this direction are certainly worthwhile.

\vskip3mm

\noindent {\bf Acknowledgments}

We would like to thank T. Azeyanagi, K. Hashimoto, K. Maeda, S.
Minwalla, M. Morikawa, T. Nishioka, T. Shiromizu and A. Strominger
for stimulating discussions.

The work of TH is supported in part by JSPS Research Fellowships for
Young Scientists. The work of SM is supported in part by MEXT
through a Grant-in-Aid for Young Scientists (B) No.~17740134, and by
JSPS through a Grant-in-Aid for Creative Scientific Research
No.~19GS0219 and through a Grant-in-Aid for Scientific Research (B)
No.~19340054. This work was also supported by World Premier
International Research Center Initiative¡ÊWPI Initiative), MEXT,
Japan. The work of TT is supported in part by JSPS Grant-in-Aid for
Scientific Research No.18840027 and by JSPS Grant-in-Aid for
Creative Scientific Research No. 19GS0219.

\vskip2mm

\appendix
\section{Solution with Electric Field}\label{solwithee}
\setcounter{equation}{0} {\hspace{5mm}}

Let us take into account the gauge field $A_{\mu}$ on the
brane and construct the exact solution for DBI action as we did in section 2. The electric flux is given by $f=F_{tx}=\de_x A_t-\de_t A_x$.
The lagrangian looks like \be L=\s{1-\dot{y}^2+(y')^2-f^2}. \ee The
equations of motion reads \be
\f{d}{dt}\left(\f{\dot{y}}{L}\right)-\f{d}{dx}\left(\f{y'}{L}\right)=0,
\ \ \ \ \ \de_x \left(\f{f}{L}\right)=0,\ \ \ \ \  \de_t
\left(\f{f}{L}\right)=0. \label{eome}\ee

Thus we can set $\f{f}{L}=c$, where $c$ is a constant. Since
$L=\f{1}{\s{1+c^2}}\s{1-\dot{y}^2+y'^2}$, the first equation in
(\ref{eome}) is the same as the one without any flux. Thus we find
that the solution is given by \ba && y=\ap\log
\cosh\f{t}{\ap}-\ap\log \cos\f{x}{\ap},\no && f=\f{c^2}{1+c^2}\cdot
\left(1-\f{\sinh^2\f{t}{\ap}}{\cosh^2\f{t}{\ap}}
+\f{\sin^2\f{x}{\ap}}{\cos^2\f{x}{\ap}}\right). \ea

\section{Details of Induced Metric on the Moving D-brane}\label{globalm}
\setcounter{equation}{0} {\hspace{5mm}} Here we present the details
of the analysis of the induced metric and the useful coordinate
transformations employed in section 3. In particular we construct a
smooth coordinate $(\ti{U},\ti{V})$ which globally covers the total
spacetime.

We start with the metric (\ref{indm}) and rewrite it into the
expression in the conformal gauge. Requiring that the coordinate $y$
is same as the one in (\ref{scherk}), this is done by performing the
following coordinate transformation \ba \tau&=&
\ap\mbox{Arctanh}\left[\f{\tan(x/\ap)\tanh(t/\ap)}{\s{1-\tanh^2(t/\ap)+\tan^2(x/\ap)}}\right]
\no &=&\ap\log
\left[\s{1+\sinh^2(t/\ap)\sin^2(x/\ap)}+\sin(x/\ap)\sinh(t/\ap)\right],
\no y&=&\ap\log \cosh(t/\ap)-\ap \log \cos(x/\ap). \label{acot} \ea
Equivalently, it is also expressed as follows \be
\sinh(\tau/\ap)=\sinh(t/\ap)\sin(x/\ap),\ \ \ \ \ \ \ \
e^{y/\ap}=\f{\cosh(t/\ap)}{\cos(x/\ap)}. \ee Note that it is also
possible to express $t$ in terms of $\tau$ by \be
\cosh^2(t/\ap)=e^{y/\ap}\left(
\cosh(y/\ap)-\s{\sinh\f{\tau+y}{\ap}\cdot\sinh\f{y-\tau}{\ap}}\right).
\ee

This coordinate transformation (\ref{acot}) leads to the metric,
which is manifestly conformally flat \be ds^2=A(\tau,y)(-d
\tau^2+dy^2), \label{aind} \ee where the conformal factor is given
by \be A(\tau,y)=1+\f{1}{\tan^2
(x/\ap)-\tanh^2(t/\ap)}=\f{1}{2}+\f{\cosh(y/\ap)}
{2\s{\sinh\f{\tau+y}{\ap}\cdot\sinh\f{y-\tau}{\ap}}}.
\label{ainduceda}\ee After we define \be u=\tau-y,\ \ \ \ v=\tau+y,
\ee the metric is expressed as (\ref{induced}).

We need to deal with the metric (\ref{induced}) carefully. It is a
good metric for $y>|\tau|$. When we go beyond this bound, the
coordinate $y$ (and $\tau$) in (\ref{induced}) becomes time-like
(space-like) and we have to exchange $\tau$ with $y$ and vise versa,
which is equal to the sign flip of $u$ and $v$. Also at the bound
$y=\pm \tau$, the metric becomes singular. Actually, we can further
perform a coordinate transformation and obtain a completely smooth
metric which covers the total spacetime on the D-string.

 This can be done by considering the following
coordinate transformation\footnote{Another way to reach this
coordinate transformation is to change the metric at $t=0$ into the
one in flat space \be ds^2=-dt^2+\f{dx^2}{\cos^2\f{x}{\ap}}\equiv
-dt^2+d\ti{y}^2, \ee where $\ti{y}=\ap\tan\f{x}{\ap}$. Then the
corresponding light-cone coordinate defined by \be \ti{U}=t-\ti{y},\
\ \ \ \ti{V}=t+\ti{y},\ee reproduced the above one (\ref{acorgl}).}
\be e^{|u|/\ap}=\cosh\f{\ti{U}}{\ap},\ \ \ \
e^{|v|/\ap}=\cosh\f{\ti{V}}{\ap}. \label{acorgl} \ee Here we took
the absolute values of $u$ and $v$ and this corresponds to the
exchange of time and space coordinate as we mentioned. Equally, \be
\ti{U}=-\ap\log\left(e^{|u|/\ap}+\s{e^{2|u|/\ap}-1}\right), \ \ \ \
\ti{V}=\ap\log\left(e^{|v|/\ap}+\s{e^{2|v|/\ap}-1}\right). \ee
 In this global
coordinate system, the metric reads \be
ds^2=-\hat{A}(\ti{U},\ti{V})d\ti{U}d\ti{V}, \label{aglobal}\ee where
\be
\hat{A}(\ti{U},\ti{V})=\f{\tanh\f{-\ti{U}}{\ap}\tanh\f{\ti{V}}{\ap}}{2}
\cdot\left(1+\f{1+\cosh\f{\ti{V}}{\ap}\cosh\f{-\ti{U}}{\ap}}
{\sinh\f{\ti{V}}{\ap}\sinh\f{-\ti{U}}{\ap}}\right). \ee  The above
metric is regular at $u=0$ and $v=0$. Thus this metric covers the
total space time and $\ti{U}$ and $\ti{V}$ take the values
$-\infty<\ti{U},\ti{V}<\infty$. In this coordinate system, the
turning point $x=0$ is mapped to $\ti{U}=\ti{V}$, while the
originally singular null lines $uv=0$ is mapped to the smooth null
lines $\ti{U}\ti{V}=0$.

The metric (\ref{aglobal}) at the null infinity $\ti{V}=\infty$
approaches \be \hat{A}(\ti{U},\ti{V})=\f{1-\tanh\f{\ti{U}}{\ap}}{2}.
\ee On the other hand in the limit $\ti{U}=-\infty$ we get \be
\hat{A}(\ti{U},\ti{V})=\f{1+\tanh\f{\ti{V}}{\ap}}{2}. \ee Thus we
can define the asymptotically flat metric by \be
U=\f{\ti{U}}{2}-\f{\ap}{2}\log\cosh\f{\ti{U}}{\ap}. \ee In the same
way we define\footnote{The inverse relation is \be
e^{-\f{2\ti{U}}{\ap}}=2e^{-\f{2U}{\ap}}-1,\ \ \ \
e^{\f{2\ti{V}}{\ap}}=2e^{\f{2V}{\ap}}-1. \ee} \be
V=\f{\ti{V}}{2}+\f{\ap}{2}\log\cosh\f{\ti{V}}{\ap}. \ee Finally,
this leads to the asymptotically flat metric (\ref{flatm}), which is
employed throughout in this paper.

\section{Exact Bogolubov Coefficients for Moving Mirror}\label{apbo}
\setcounter{equation}{0} {\hspace{5mm} } Consider a perfect
reflecting mirror is placed in a $(1+1)$ dimensional massless scalar
field theory. Following our setup (\ref{eqn:P(U)}), we assume that
its trajectory is described by $y=\ti{\ap}\log\cosh\f{t}{\ti{\ap}}$,
where we rescaled the parameter $\ap$ in (\ref{curve}) by factor two
as $\ti{\ap}=\f{\ap}{2}$. We define the following function $p(u)$
($u=t-y,~ v=t+y$) as in the standard argument e.g.
\cite{Birrell:1982ix}

\begin{equation}
p(u)= - \ti{\ap} \ln (2- e^{u/\ti{\ap}}),
\end{equation}
or
\begin{equation}
p'(u)= \frac{e^{u/\ti{\ap}}}{ (2- e^{u/\ti{\ap}})}.
\end{equation}
We have the following two choices of normalized outgoing modes:
\begin{eqnarray}
\phi^{in}_ \omega &=& \frac{i}{\sqrt{4 \pi \omega} } e^{-i \omega p(u)}\\
\phi^{out}_ {\omega'} &=& \frac{i}{\sqrt {4 \pi \omega'} } e^{-i
\omega' u},
\end{eqnarray}
where $\phi^{in}$ corresponds to the reflection of a standard
incoming wave, while $\phi^{out}$ to the standard outgoing mode. In
other words, the vacuum with respect to $\phi^{in}$ (or
$\phi^{out}$) defines the one for in vacuum (or out vacuum).

Then the Bogolubov coefficient is expressed as follows \be
\ap_{\omega, \omega'}= (\phi^{in}_\omega ,\phi^{out}_{\omega'}  ),\
\ \ \ \beta_{\omega, \omega'}= (\phi^{in}_\omega
,\ov{\phi^{out}_{\omega'}} ). \ee More explicitly, for example
\begin{eqnarray}
\ap_{\omega, \omega'} &=& -i \int_{t- fixed} dy  (\phi^{in}_\omega
\partial _t \phi^{out}_{\omega'}
- \partial _t \phi^{in}_\omega   \phi^{out}_{\omega'} )\\
&=& i \int_{- \infty} ^ {\ti{\ap} \ln 2} du ( \phi^{in}_\omega
\partial _u \phi^{out}_{\omega'}
 - \partial _u \phi^{in}_\omega   \phi^{out}_{\omega'}).
\end{eqnarray}
Using the formula like
\begin{eqnarray}
&&\int^{\ti{\ap} \ln 2}_{- \infty} (2- e^{u/ \ti{\ap}})^{i \omega
\ti{\ap}} e^{i \omega 'u} =-i 2 ^{i (\omega+ \omega') \ti{\ap} }
 \frac{\Gamma(1+ i \omega \ti{\ap})\Gamma(1+ i \omega' \ti{\ap})}{\omega '\Gamma(1+ i \omega \ti{\ap}
 + i \omega' \ti{\ap})},\no
&&\int^{\ti{\ap} \ln 2}_{- \infty} (2- e^{u/ \ti{\ap}})^{i \omega
\ti{\ap}-1} e^{i \omega 'u+ u/ \ti{\ap}} =- i 2 ^{i (\omega+
\omega') \ti{\ap} }
 \frac{ \Gamma( 1+i \omega \ti{\ap})\Gamma(1+ i \omega' \ti{\ap})}{\omega \Gamma(1+ i \omega
 \ti{\ap}+ i \omega' \ti{\ap})},
\end{eqnarray}
in the end, we find the exact Bogolubov coefficients $\ap_{\omega,
\omega'}$ and $\beta_{\omega, \omega'}$ \ba &&\ap_{\omega, \omega'}
 =\frac{i}{2 \pi\sqrt{\omega \omega'}} 2 ^{i (\omega+ \omega') \ti{\ap} }
 \frac{ \Gamma( 1+i \omega \ti{\ap})\Gamma(1+ i \omega' \ti{\ap})}
 { \Gamma(1+ i \omega \ti{\ap}+ i \omega' \ti{\ap})}\no
 && \beta_{\omega,\omega'}
 =-\frac{i}{2 \pi\sqrt{\omega \omega'}} 2 ^{i (\omega- \omega') \ti{\ap} }
 \frac{ \Gamma( 1+i \omega \ti{\ap})\Gamma(1- i \omega' \ti{\ap})}
 { \Gamma(1+ i \omega \ti{\ap}- i \omega' \ti{\ap})}.
\ea Their absolute value squares are
\begin{eqnarray}
|\ap_{\omega,\omega'}|^2&=& \frac{\ti{\ap}}{2\pi(\omega + \omega')}
\frac{(e^{2 \ti{\ap} \pi (\omega + \omega') }-1)} {(e^{2
\ti{\ap} \pi \omega }-1)(e^{2 \ti{\ap} \pi \omega' }-1)}\\
|\beta_{\omega,\omega'}|^2&=& \frac{\ti{\ap}}{2\pi(\omega -\omega')}
\frac{(e^{2 \ti{\ap} \pi (\omega - \omega') }-1)} {(e^{2 \ti{\ap}
\pi \omega }-1)(1-e^{-2 \ti{\ap} \pi \omega' })}.
\end{eqnarray}

The total creation of the particle with the energy $\omega$ is given
by \be N_\omega=\int_0^\infty d\omega'|\beta_{\omega,\omega'}|^2\sim
\f{A}{e^{2\pi\ti{\ap}\omega}-1}, \ee where $A$ is a divergent
constant\footnote{This divergence is also natural from the constant
energy flux (\ref{efm}). If we integrates it for an infinitely long
time, we encounter the same divergence.}. This is because
$|\beta_{\omega,\omega'}|^2$ behaves $\sim\f{\ti{\ap}}{2\pi\omega'}
\f{1}{e^{2\pi\ti{\ap}\omega}-1}$ when $\omega'$ is very large.
Therefore the particle creation leads to a thermal spectrum with the
temperature $T=\f{1}{2\pi\ti{\ap}}$.

\section{Fluctuations on Curved D-brane}\label{DBIa}
\setcounter{equation}{0} {\hspace{5mm}} Let us consider an $n$
dimensional timelike hypersurface $\Sigma$ in an ($n+1$)-dimensional
spacetime. Provided that $\Sigma$ is sufficiently smooth, it is
always possible to adopt a Gaussian normal coordinate system in a
neighborhood of $\Sigma$ so that the spacetime metric is of the form
%
\begin{equation}
 ds^2 = \tilde{g}_{\mu\nu}dx^{\mu}dx^{\nu} + dw^2,
  \label{eqn:n+1decomposition}
\end{equation}
where $x^{\mu}$ ($\mu=0,\cdots,n-1$)  are intrinsic coordinates on
$\Sigma$ and $w$ is the geodesic distance from $\Sigma$. Note that
$\tilde{g}_{\mu\nu}$  depends not only on $x^{\mu}$  but also on $w$. In
this coordinate system the hypersurface $\Sigma$ is represented as $w=0$
and the induced metric on $\Sigma$ is
$g_{\mu\nu}\equiv\tilde{g}_{\mu\nu}|_{w=0}$.

With the bulk metric and the above Gaussian normal coordinate system
fixed, we now consider a slightly perturbed hypersurface $\Sigma'$
represented as $w=\phi(x)$. The induced metric on the perturbed
hypersurface $\Sigma'$ is
%
\begin{equation}
 g'_{\mu\nu} =
  \left.(\tilde{g}_{\mu\nu} +
   \partial_{\mu}\phi\partial_{\nu}\phi)\right|_{w=\phi(x)}.
\end{equation}
By expanding this with respect to $\phi$ and its derivatives up to
second order, we obtain
%
\begin{equation}
 g'_{\mu\nu} = g_{\mu\nu} + h_{\mu\nu},  \quad
 h_{\mu\nu} \equiv 2K_{\mu\nu}\phi + K'_{\mu\nu}\phi^2
  + \partial_{\mu}\phi\partial_{\nu}\phi + O(\phi^3),
  \label{eqn:hmunu}
\end{equation}
where
$K_{\mu\nu}\equiv \frac{1}{2}\partial_w\tilde{g}_{\mu\nu}|_{w=0}$ and
$K'_{\mu\nu}\equiv \frac{1}{2}\partial_w^2\tilde{g}_{\mu\nu}|_{w=0}$.

Expanding $\sqrt{-g'}$ with respect to $h_{\mu\nu}$ and keeping terms up
to the second order, we obtain
%
\begin{equation}
 \sqrt{-g'} = \sqrt{-g}
  \left\{ 1+\frac{1}{2}h
   +\frac{1}{2}\left(\frac{1}{4}h^2-\frac{1}{2}h^{\mu\nu}h_{\mu\nu}\right)
   + O(h^3)\right\},
\end{equation}
where $h\equiv g^{\mu\nu}h_{\mu\nu}$ and
$h^{\mu\nu}\equiv g^{\mu\rho}g^{\nu\sigma}h_{\rho\sigma}$. By
substituting the expression (\ref{eqn:hmunu}), we obtain
%
\begin{equation}
 \sqrt{-g'} = \sqrt{-g}
  \left\{ 1 + K\phi +
   \frac{1}{2}g^{\mu\nu}\partial_{\mu}\phi\partial_{\nu}\phi
   + \frac{1}{2}m^2\phi^2 + O(\phi^3)\right\},
\end{equation}
where
%
\begin{equation}
 m^2 \equiv g^{\mu\nu}K'_{\mu\nu} + K^2 - 2K^{\mu\nu}K_{\mu\nu},
  \label{eqn:m2}
\end{equation}
$K\equiv g^{\mu\nu}K_{\mu\nu}$ and
$K^{\mu\nu}\equiv g^{\mu\rho}g^{\nu\sigma}K_{\rho\sigma}$.

Now, corresponding to the ($n+1$)-decomposition of the bulk metric
(\ref{eqn:n+1decomposition}), the Ricci scalar at $w=0$ is decomposed as
%
\begin{equation}
 \left.R^{(n+1)}\right|_{w=0}
  = R^{(n)} - 2g^{\mu\nu}K'_{\mu\nu} - K^2 + 3K^{\mu\nu}K_{\mu\nu},
\end{equation}
where $R^{(n)}$ is the Ricci scalar of $g_{\mu\nu}$. Thus, $m^2$ in
(\ref{eqn:m2}) is rewritten as
%
\begin{equation}
 m^2 =
  \frac{1}{2}\left\{R^{(n)} - \left.R^{(n+1)}\right|_{w=0} + K^2
          - K^{\mu\nu}K_{\mu\nu}\right\}.
\end{equation}
If the unperturbed hypersurface $\Sigma$ is a minimal surface
satisfying $K=0$ and if $\left.R^{(n+1)}\right|_{w=0}=0$ then this
formula is reduced to
%
\begin{equation}
 m^2 =
  \frac{1}{2}(R^{(n)} - K^{\mu\nu}K_{\mu\nu}).
\end{equation}

\end{document}